# Chapter 4: Astrochemistry: Synthesis and Modelling

**Valentine Wakelam**

Univ. Bordeaux, LAB, UMR 5804, F-33270, Floirac, France

CNRS, LAB, UMR 5804, F-33270, Floirac, France

**Herma M. Cuppen**

Theoretical Chemistry, Institute for Molecules and Materials, Radboud University Nijmegen, Heyendaalseweg 135, 6525 AJ Nijmegen, The Netherlands

**Eric Herbst**

Departments of Chemistry, Astronomy, and Physics, University of Virginia, Charlottesville, VA 22904 USA

**Abstract**   We discuss models that astrochemists have developed to study the chemical composition of the interstellar medium. These models aim at computing the evolution of the chemical composition of a mixture of gas and dust under astrophysical conditions. These conditions, as well as the geometry and the physical dynamics, have to be adapted to the objects being studied because different classes of objects have very different characteristics (temperatures, densities, UV radiation fields, geometry, history etc); e.g., proto-planetary disks do not have the same characteristics as protostellar envelopes. Chemical models are being improved continually thanks to comparisons with observations but also thanks to laboratory and theoretical work in which the individual processes are studied.

## 4.1 Introduction

A large number of molecules have now been observed in the interstellar medium (ISM) and many more are expected to be discovered considering the unidentified lines in existing spectral surveys and new surveys to come [1]. The promise of **ALMA**[1], a powerful new interferometric telescope, on this matter is endless. On a very basic level, simple molecules such as CO and CS are used to understand the physical properties of astrophysical objects such as **dark clouds**, **star forming regions**, **proto-planetary disks,** and galaxies through the excitation of their observed spectral lines [2] (see also chapter XXX, REFERENCE TO ANOTHER CHAPTER). Observers need, however, to know the distribution of these and other species in terms of **abundances**[2] in these sources. Chemical models are used for this purpose; e.g., to understand the chemical composition of astrophysical objects and their evolution. Combining model predictions with radiative transfer, astrochemists can make predictions on the detectability of some species. In addition, the abundance of "key" species that cannot be directly observed (molecules without a **dipole moment** for instance) can be computed with these models. These computations can be important for numerical simulations of protostellar collapse, for instance, since the energy budget (cooling and heating) depends on the abundances of some cooling species such as atomic oxygen that cannot be directly observed except in exceptional regions.

The gas and dust in the ISM are constantly being refreshed and modified, a process that is related to the life cycle of stars (reference to another chapter??). After the death of a star, the elements that are formed inside the star are spread throughout the diffuse interstellar medium in the form of atoms and refractory dust particles, often known as "grains." This material is modified somewhat by chemical processes and by the strong interstellar UV field at play in such environments, before gravity collapses the material to form colder dense clouds, where external UV photons cannot penetrate. In the interior of these dense clouds, a rich gasphase chemistry takes place and the gas interacts with grain surfaces where catalytic surface reactions can occur as well. Species such as molecular ions, radicals, isomers, and large linear unsaturated molecules such as $HC_{11}N$ can then be formed in the gas, while more saturated species such as water, methane, ammonia, and methanol can be formed on the grains along with $CO_2$. When stars and planets form from these clouds, this molecular complexity mutates into a more terrestrial-like organic chemistry including molecules such as basic esters, alcohols, and more saturated nitriles, which can survive and participate in the very long path towards life.

---

[1] ALMA, for Atacama Large Millimetre Array: https://science.nrao.edu/facilities/alma

[2] Abundances are defined by the ratio of the density of a species to the total density (in cm$^{-3}$) of hydrogen atoms. Because hydrogen is mostly in the form of atomic H or molecular $H_2$, the proton density is $n_H = n(H) + 2n(H_2)$.

To study how the astrophysical environment modifies the composition of the gas and the dust, astrochemists have built increasingly complex chemical models over the years. In this chapter, we will describe those models (section 4.2), discuss the chemical and physical processes that occur in the ISM (section 4.3), and consider the synthesis of $O_2$ as an example of how these processes operate (section 4.4).

## 4.2 Astrochemical models

From a numerical point of view, chemical models solve a system of differential equations of the type:

$$\frac{dn_i}{dt} = \sum production - \sum destruction \qquad (4.1)$$

where $n_i$ is the density of species $i$ (in cm$^{-3}$). The production and destruction terms refer to all chemical and physical processes that produce and destroy this species. Numerical solvers using the Gear method [3] are typically employed in double precision to solve such systems as a function of time. In the end, modellers compute the evolution of the densities of species (or abundances) for a set of parameters and from an initial composition. For large systems, the number of equations (one for each species) and processes can be quite important and lead to hypersensitivity to some model parameters and even bi-stability[3] [4,5].

In fig.4.1, we show the results of a model for dense cloud conditions (a temperature of 10 K, a hydrogen atom density of $2x10^4$ cm$^{-3}$ and a **visual extinction** of 10), which includes only gas-phase processes except for the production of molecular hydrogen, which occurs on granular surfaces. The chemical processes that are involved in the chemistry of the **interstellar medium** are described in section 4.3 of this chapter. In addition to the parameters described in section 4.2.1, the geometry of the object, the presence of **mixing,** and physical dynamics can influence the chemical composition as well (sections 4.2.2 and 4.2.3).

---

[3] Bi-stability refers to two different steady-state results with the same set of parameters but different initial conditions.

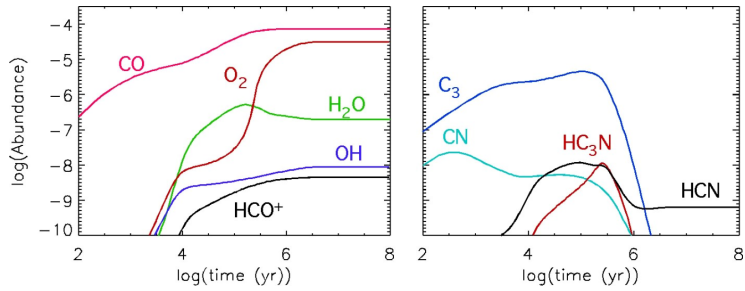

**Fig. 4.1** Abundances (compared to total H) of species computed by a pure gas-phase chemical model for dense cloud conditions as a function of time.

### 4.2.1 Elemental abundances and initial conditions

Material consisting of various elements, related to the life cycle of stars, is constantly modifying the existing material in the ISM. The elements are eventually stored in three forms [6]:

- *phase 1*: in the gas-phase (in atomic or molecular form),
- *phase 2*: in refractory cores of interstellar grains (formed in circumstellar envelopes),
- *phase 3*: in a mantle of volatile species on top of grain cores.

The formation of grain mantles is efficient in cold regions shielded from UV photons. **Depletion** of the elements from purely gas-phase systems first occurs in those parts of stellar atmospheres where grains are first formed (see fig.4.3). Additional depletion from the gas-phase has been observed in diffuse clouds depending on the overall density of the clouds (from a few atoms and molecules per $cm^3$ to 100 per $cm^3$) [7,8]. Although it seems logical to assume that the missing elements stick on grains, the mechanisms of depletion (considering the very low densities of these clouds) and the form of the depleted species remain a puzzle [9,10]. Observations of unambiguous gas-phase elemental abundances are limited to purely atomic diffuse clouds with densities smaller than 10 $cm^{-3}$, where the UV field is still very efficient in preventing the formation of molecules in the gas-phase and at the surface of the grains. As a consequence, we do not have direct measurements of the fraction of the elements that are available for the formation of the molecules observed in dense clouds (in the gas and in the grain mantles). Most (if not all) chemical models of dense sources (molecular clouds, proto-stellar envelopes, proto-planetary disks) include as initial input an additional depletion of those elements heavier than oxygen compared with the gas-phase abundances observed in diffuse clouds, which are already depleted in these elements compared with stellar abundances, except for isolated exceptions such as sulphur. Many studies even consider the elemental abundances as free parameters that can be varied in order to reproduce the observations of gas-phase molecules in dense regions, even for elements lighter than oxygen such as nitrogen. Nitrogen (for example [11,12]) and sulphur (for example [13,14]) are good examples in this respect. Even when they are not varied intentionally, uncertainties in their exact values remain, depending on the reference used [15], and the results of chemical models strongly depend on these choices [16].

Another type of parameter, to which the model can be sensitive, is the choice in the initial composition of species (initial conditions). Since the chemistry is not at steady-state in most objects, the chemical composition predicted by the model will depend on the assumed initial conditions. There does not exist yet any model able to follow the chemical composition of the gas and dust during a complete cycle of evolution starting with material ejected from stars and ending with the collapse of clouds to form new stars, mainly because the evolution between different stages of star formation (e.g., diffuse to dense clouds, proto-stellar envelopes to proto-planetary disks) is not fully understood. For these reasons, assumptions about previous steps are usually made. For dense clouds for instance, it is typically assumed that they are formed from diffuse medium conditions in which elements are mainly in the atomic form and that the time scale of the "contraction" up to the densities of dense clouds is faster than the chemical evolution. Some models do explicitly include the physical transition from one stage to another as the chemistry progresses. As an example, Hassel et al. [17] followed the chemistry as a shock wave passes through diffuse material and a cold dense cloud begins to form behind the shock. Unfortunately, the abundances of gas-phase and grain-surface molecules synthesised were only followed up to a **visual extinction** of 3, at densities not as high as are found in cold dense clouds, because of limitations to the physical model. For proto-planetary disks, the chemical composition of the parent cloud is usually assumed to define the initial conditions [18]. In pursuing such an approximation, we of course ignore the transition between the cloud and the disk itself and so assume that the chemical composition is not modified during this transition. In the absence of a full physical model describing those transitions, it is probably the best that can be done.

## 4.2.2 Geometry

In astrophysical sources, where turbulence, or any kind of mixing, is not efficient (or is not constrained), the chemistry is usually treated in a zero-dimensional ap-proximation, which consists of single spatial point. In such models, the temperature and densities are kept constant in space and time and the chemical composition is computed at each time step from an initial composition up to steady-state (or earlier if the time to reach steady-state is unrealistic). Many (if not all) astrophysical objects cannot be characterised by a single temperature and density, but present spatial gradients. Envelopes of gas and dust around proto-stars are characterised by an increase of the temperature and density towards the centre of the envelope in a spherical symmetry. Figure 4.2 shows the temperature and density in the envelope of the low mass proto-star IRAS16293-2422 as determined by Crimier et al. [19] based on the analysis of observations of the dust and the $H_2O$ molecule at several wavelengths. In such objects, the chemistry will strongly depend on the radius. At high temperatures (larger than the specific sublimation temperature of the species), many molecules from the grain mantles will be evaporated in the gas-phase and will participate in the gas-phase chemistry (see for instance [20,21]). Such an object (if considered as static) can be treated in a so-called pseudo 1D approach; i.e. the temperature and density will depend on the radius but the cells are independent. There is observational evidence from non-thermal broadening of the molecular lines that turbulent mixing exists in these objects although the exact nature of the mixing is not understood. Such mixing processes can be included in chemical composition calculations, as was done by Wakelam et al. [22] for massive proto-stellar envelopes. In proto-planetary disks, where radial and vertical mixing may exist, 2D chemical models with mixing have been developed [23].

The geometry is also important for the treatment of the interaction with UV photons. Borders of dense molecular clouds are exposed to the interstellar UV field produced by massive stars. The penetration of these photons into the cloud has to be computed as a function of depth since they are absorbed and scattered by the dust (see section 4.3.1.1). In proto-planetary disks, the young central star usually presents a strong emission of UV (and X-ray) photons so that the UV penetration has to be computed in two dimensions to take into account the interstellar UV radiation field and the one coming from the star.

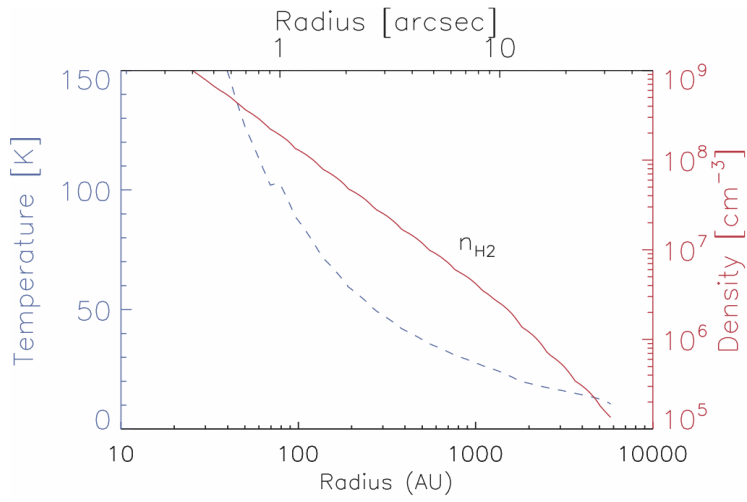

**Fig. 4.2** Temperature and density of H$_2$ profiles in the envelope of the low mass proto-star IRAS 16293-2422 based on multi-wavelength observations from Crimier et al. [19].

### 4.2.3 Physical conditions evolving with time (dynamics)

Most, if not all, astrophysical objects are not static over a period of time long enough for the chemistry to reach steady state, except perhaps in the diffuse medium as long as the initial H$_2$/H abundance ratio is assumed to be non-zero. As a consequence, the modifications of the physical conditions with time have to be considered. A good example is the modelling of **circumstellar envelopes.** The cells of material pushed away from the star encounter lower temperatures and densities. Close to the star, the temperatures are so high that species are only in atomic form. As the material moves away from the star, molecules will be formed and survive until they encounter a much thinner medium where the interstellar UV field will destroy them. A schematic view of the physical structure, which is far more complex than discussed here, and the chemical composition of a circumstellar envelope are given by Fig. 4.3.

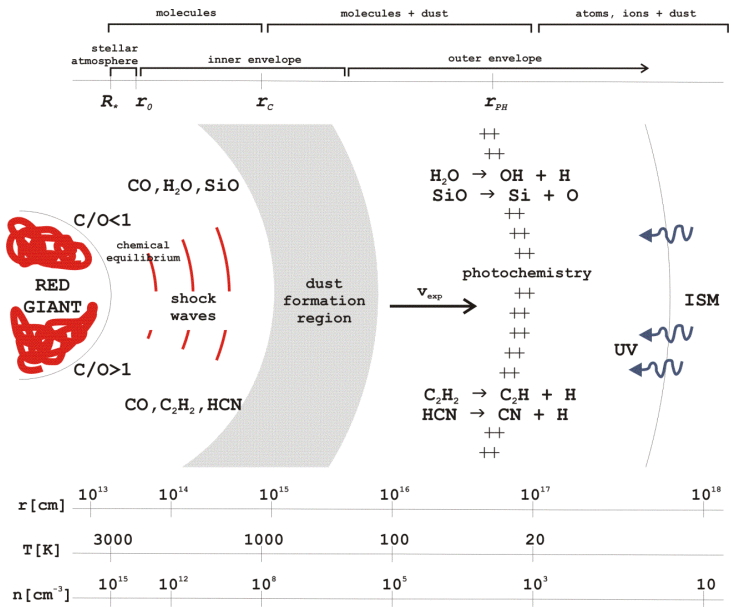

**Fig. 4.3** Representation of a circumstellar envelope (courtesy of Marcelino Agúndez).

As another example, consider the formation of stars, which begins with the collapse of a molecular cloud first into a pre-stellar core and then into a proto-star. This collapse occurs in a few times $10^5$ yr, during which the temperature and the density in the proto-stellar envelope will increase towards the centre. The environment is often referred to as a hot core or, for low-mass stars, a hot corino, when the temperature reaches $100 - 300$ K [24]. During this time, the material of the envelope can fall towards the newly forming star in a rotating motion. The gas and dust will first encounter a small temperature and large density allowing most of the molecules to stick onto the grains, then a medium temperature during which grain-surface reactions will take place, and then a warmer temperature allowing the sublimation of the molecules formed on the grains. Although simplified models exist in which the time-dependent temperature is homogeneous throughout the warming region [25], a more accurate approach is to couple the chemistry to a 1D hydrodynamic collapse model, which computes the evolution of the physical conditions in a Lagrangian approach [21]. An extension to a full 3D hydrodynamic model is underway [26].

## 4.3. Physico-chemical processes

The chemistry of the ISM can be separated into four groups of processes. The first is the interaction with high-energy cosmic-ray particles, while the second consists of photo-processes induced by UV photons[4]. The third concerns bimolecular gas-phase reactions, and the last one concerns interactions with grains. A summary of the dominant processes in a selection of astrophysical objects is given in fig. 4.4. We discuss the first three classes in the section 4.3.1 and the last one in section 4.3.2.

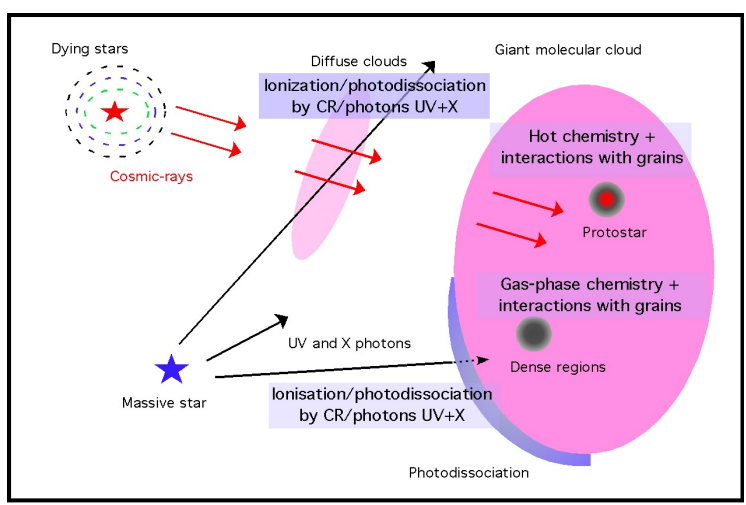

**Fig. 4.4** Schematic view of the dominant physico-chemical processes in the interstellar medium depending on the physical conditions.

### 4.3.1 Gas-phase chemistry

### 4.3.1.1 Processes and parameters

---

[4] X-rays are also important in some sources.

Cosmic-ray particles (with energies in the MeV to GeV range) produce the ionisation of molecular and atomic hydrogen and helium. The high-energy electrons from this process can in turn excite $H_2$, which then emits UV photons. This process, first proposed by Prasad & Tarafdar [27], is an efficient source for the photodissociation of molecules in the dense internal part of molecular clouds. Although these first-order processes are not standard bimolecular chemical processes, their rate can be expressed in term of rate coefficients. The rates of direct and indirect dissociation and ionisation by cosmic rays are proportional to the total $H_2$ cosmic-ray ionisation rate $\zeta$ [28,29]. As an example, the helium direct cosmic-ray ionisation rate is one-half of $\zeta$. The value of $\zeta$ for any species is a function of depth into an interstellar cloud, although this dependence is most frequently ignored because it is difficult to calculate [30].

The rates of photodissociation caused by external UV photons depend on the **visual extinction** $A_V$:

$$k_{phot} = a\exp(-\gamma A_V) \qquad (4.2)$$

with $\alpha$ and $\gamma$ parameters specific to each species. The visual extinction $A_V$ is proportional to the total hydrogen **column density** $N_H$ ($A_V = N_H / 1.6\times10^{21}$, [31]) where column density ($cm^{-2}$) is defined as the volume density integrated over a path through the cloud towards the observer.

Bimolecular reactions have largely been described in Chapter 1 of this book. The efficiency of these processes can depend on the temperature of the gas. All production and destruction terms (in units of $cm^{-3}$ $s^{-1}$) in equation (1) can be written as $k_{ij}n_in_j$ for bimolecular (second-order) reactions and $k_in_i$ for first-order reactions, with $k_{ij}$ the rate coefficient of the reaction between species $i$ and $j$ and $n$ the density of the reactant(s). More details on these expressions can be found in Wakelam et al. [22].

Current networks for astrochemical models contain more than four thousand gas-phase reactions for more than four hundred atomic and molecular species, which comprise neutral, positively, and negatively charged species [21]. Most of these reactions have not been studied under the conditions of the cold ISM so that the rate coefficients can be quite uncertain if not completely wrong [16]. An example of the latter concerns the formation of OCS in the gas-phase by the radiative association reaction CO + S → OCS + hv. A rate coefficient of $1.6\times10^{-17}$ $(T/300)^{-1.5}$ $cm^3$ $s^{-1}$ was included in models following a crude estimate by Prasad & Huntress [28], which gives a rate coefficient of approximately $2\times10^{-15}$ $cm^3$ $s^{-1}$ at 10 K. Recent calculations by Loison et al. [32] show that the rate was largely overestimated at 10 K. The consequence for the OCS gas-phase production is then dramatic and the predicted abundance of OCS much smaller than the observed one. In general, radiative association reactions, despite their importance, have not been studied in more than a few cases in the laboratory at any temperature.

### 4.3.1.2 Uncertainties and sensitivity to the parameters

Model predictions can be very sensitive to model parameters. Uncertainties in the parameters can therefore lead to strong differences in the predicted abundances of species. Two aspects of the problem can be considered. First, the model parameters are more or less known within a range of uncertainty. These uncertainties propagate through the time-dependent calculations and lead to error bars for the model results. If those error bars have been derived, quantitative comparisons with observations (for which error bars are more frequently determined) can be undertaken. Secondly, it is possible through sensitivity analyses to identify key parameters for which a better estimate would reduce the model uncertainties. A number of such studies have been done recently with the aim of 1) computing model error bars, 2) understanding the sensitivity of these models to the parameters (sensitivity analysis), and 3) identifying key parameters in a variety of sources.

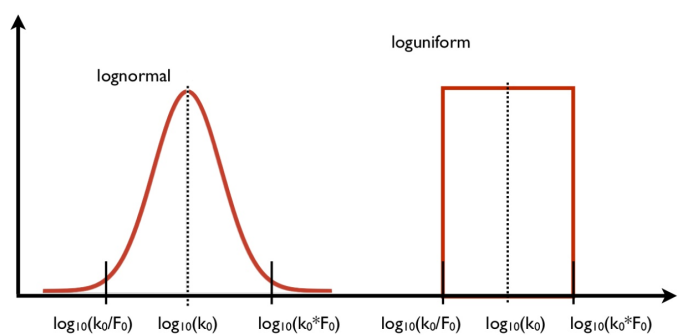

**Fig. 4.5** Possible distributions (lognormal or loguniform) of the rate coefficients randomly varied within a factor $F_0$.

Estimation of model error bars and sensitivity analyses are based on the same principle. All rate coefficients (or other model parameters) of a system are randomly varied within a certain range. The chemical evolution is then computed for each set of rate coefficients. For a network containing four thousand reactions, the model is typically run two thousand times with different sets of rate coefficients. The distribution of the rate coefficients can be either log-normal or log-uniform (see fig. 4.5). The first choice implies that the mean value $k_0$ is a preferred value. This is usually the case for rate coefficients, which are measured with an uncertainty defined by statistical errors. The factor $F_0$, which defines the range of variation, can be a fixed factor for all reactions for a **sensitivity analysis** or specific to each reaction for an uncertainty propagation study. Use of the same $F_0$ for all reactions, in the case of a sensitivity analysis, assures the modeller that an underesti-mated uncertainty factor will not bias the analysis. The results of thousands of runs are used differently to identify important reactions and to estimate model error bars.

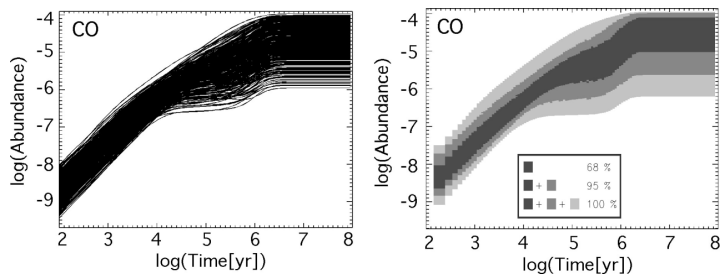

**Fig. 4.6** CO abundance computed by a chemical model under dense cloud conditions (from [16]). Each curve on the left represents the result of one model in which the rate coefficients have been randomly modified. The grey levels on the right show the density of the curves (in percentage).

For sensitivity analyses, one of the methods to identify "key"[5] reactions is to compute Pearson correlation coefficients for each species and each reaction. Such coefficients quantify how modified abundances of species correlate with the modifications of rate coefficients. For uncertain propagation, model error bars, or uncertainties in the computed abundances of species, are defined by the minimum interval containing 95% of the curves (see fig. 4.6 and [33]). When the abundances follow a log-normal distribution, these error bars are equivalent to twice the standard deviation of the curves. More details on these methods and their applications can be found in a review paper by Wakelam et al. [34].

### 4.3.1.3 Reduction of chemical networks

Large chemical networks are often needed to synthesise large molecules. For some applications, for example, in which the chemistry is calculated with a physical model that leads to lengthy computational times, it can be convenient or even necessary to reduce the size of a network of reactions, if only limited information is required. The simpler technique to accomplish this goal is to remove reactions one-at-a-time and check that the abundances of the species of interest remain within a certain range of tolerance. This of course can only be correct for one physical condition and one initial network. For the reduced network to be used over a range of physical conditions, the analysis has to be redone several times. In addition, if the rate coefficients of some reactions have to be changed, the out-come may be changed. More sophisticated methods based on correlations can be used, as they are probably less time consuming [35].

---

[5] By "key" reactions, we mean reactions with rate coefficients that are quantitatively important for the computed abundance of species.

Such reduction methods have been applied by a number of groups for different cases. To the best of our knowledge, Ruffle et al. [35] were the first to apply such a technique (in a more complex version) to determine the minimum network for the computation of the gas-phase CO abundance in interstellar clouds of different density, temperature and visual extinction. Their network still contains more than two hundred reactions and more than sixty species. Reduced networks to compute the ionisation fraction in dense clouds [36] and in proto-planetary disks [37], and reduced networks for dense clouds [38] have also been proposed.

### 4.3.2 Gas-grain interactions and surface reactions

In the ISM, molecules from the gas phase interact with grains by sticking to their surface, possibly undergoing reactions with other absorbates, and desorbing again through different thermal and non-thermal processes. The formation of molecular hydrogen, the most abundant molecule in the ISM, is the most obvious example for the importance of grain surface chemistry, since it is almost entirely formed through surface chemistry in cold environments. As explained in Chapter 1, species can either be **physisorbed** to the surface and then general reactions occur through the diffusive **Langmuir-Hinshelwood mechanism**, or be chemisorbed to the surface so that chemistry can also proceed through the **Eley-Rideal mechanism**, in which the adsorbate is struck by a gas-phase reactant. Molecular hydrogen is formed through both mechanisms; the Langmuir-Hinshelwood mechanism dominates at low temperature for both silicate and carbonaceous grains and the Eley-Rideal mechanism at high gas and/or grain temperatures for carbonaceous grains.

In the present section, we will present the different methods to model grain surface chemistry, the input parameters that go into these models, the experimental and theoretical methods to obtain these parameters, and the limitations of these models. Most gas-grain networks are predominantly designed for the low temperature regime and, as we will see in the following sections, the expressions that are used almost exclusively describe the Langmuir-Hinshelwood type of reaction.

#### 4.3.2.1 Rate equation approximation

The most obvious way to introduce grain-surface chemistry into gas-phase astrochemical models is to use rate equations for both the surface chemistry and for gas-phase chemistry. The main advantages of this method are that the gas phase and grain surface can be easily coupled and that the rate equation method is computationally light, which allows for a large number of grain-surface species and reactions. Moreover, a long timescale can be simulated and simulating a large number of points to cover a grid of different conditions is feasible as well.

Here we follow the rate equation approach by Hasegawa et al. [39]. Consider a dust particle with $N$ surface sites, on which there are $N(A)$ molecules of species A. The surface abundance, $N(A)$, changes in time according to

$$\frac{dN(A)}{dt} = k_{acc}\, n(A) - k_{des} N(A) - \sum (k_{hop,A} + k_{hop,i}) N(A) N(i) / N$$
$$+ \sum (k_{hop,i_1} + k_{hop,i_2}) N(i_1) N(i_2) / N \qquad (4.3)$$

where $n(A)$ is the gas phase abundance of species $A$ and the different $k$'s represent different rate coefficients. The first term accounts for accretion of $A$ from the gas phase to the surface; the accretion rate coefficient $k_{acc}$ is determined by the product of the velocity of $A$ in the gas phase, the cross section of the grain, and the sticking fraction of $A$ to the grain, which depends on the gas and surface temperatures. The second term represents the loss of $A$ from the surface due to desorption. This can occur through a non-thermal desorption mechanism or through thermal desorption, for which the rate coefficient is given by

$$k_{des,therm} = \nu \exp\left(-E_{des} / k_B T_{grain}\right) \qquad (4.4)$$

with $\nu$ the attempt frequency, $E_{des}$ the desorption energy, $k_B$ the Boltzmann constant, and $T_{grain}$ the grain temperature. Non-thermal mechanisms include **photodesorption** [40,41], cosmic-ray desorption due to sputtering or flash heating of the grain [42,43], and desorption upon reaction [44].

The third term in Eq. 4.3 refers to the loss of species $A$ due to Langmuir-Hinshelwood reactions with species $i$. The rate coefficient $k_{hop,i}$, defined as the rate of hopping of species $i$ over the **potential barrier** between two adjacent sites, is itself given by the equation:

$$k_{hop,i} = \nu_0 \exp\left(-E_{b,i} / k_B T_{grain}\right) \qquad (4.5)$$

Similarly the last term represents the gain in species $A$ by formation mechanisms.

Apart from the clear advantages of this method, which lies mostly in its user-friendliness and the fact that it is computationally inexpensive, there are several disadvantages to the rate equation method, which become apparent in different physical environments. In conditions where the number of species on the surface is small (<<1), one can run into the so-called "accretion" limit [45] in which the formation rate of molecules can be overestimated by several orders of magnitude. The quantities $N(i)$ are in fact expectation values of $N(i)$: $\langle N(i) \rangle$. The Langmuir-Hinshelwood term in Eq. 4.3 then becomes $\langle N(A) \rangle \langle N(i) \rangle$, where in fact the rate of reaction is determined by $\langle N(A)N(i) \rangle$. For small values of $N(A)$, this approximation is not valid and $\langle N(A) \rangle \langle N(i) \rangle$ overestimates $\langle N(A)N(i) \rangle$. Different methods have been developed to overcome this problem.

Another disadvantage of the rate equation method is that the actual surface and the positions of the atoms and molecules are not considered. Especially in conditions where the grain mantle is beyond the sub-monolayer regime, this can have severe consequences: species are allowed to interact with all other species, regardless of their relative position, species are assumed to hop and desorb with the same rate, regardless of their position (bulk vs. surface), and the grain mantle is assumed to be homogeneous, whereas observations show that the grain mantle consists of several layers of different composition. But even in the limit where the surface species add up to less than a monolayer, the rate equation method can overestimate formation rates due to the missing back diffusion term. When an atom scans the grain surface, it can visit a site more than once, which is called "back diffusion" and can lower the rate by as much as a factor of 3 [46].

### 4.3.2.2 Methods to circumvent the accretion limit

Different stochastic methods have been considered to overcome the accretion limit problem. The main focus has been on the master equation method [47,48,49] and macroscopic Monte Carlo simulations [50,51,49]. Of the two, the master equation method can be more easily coupled to rate equations, which handle the gas phase chemistry in the most straightforward way, since it treats the gas and surface kinetics with equations of similar form.

In the Monte Carlo methods, the evolution of (discrete) number densities is followed in time by randomly selecting a sequence of processes. The probability of selecting a process is proportional to its rate, which is determined in a similar manner to the rate equation method. Because Monte Carlo methods use random numbers and probabilities instead of analytical expressions, coupling between the two methods (rate equations for the gas and a Monte Carlo procedure for the grain) is harder to achieve. Different (kinetic) Monte Carlo implementations are used and usually a distinction between macroscopic and microscopic Monte Carlo is made. In the macroscopic simulations, only the number density is followed in time; in the microscopic simulations the exact positions of the species are also considered. Recently, macroscopic Monte Carlo simulations of both the gas phase and grain surface chemistry have been carried out for a proto-planetary disk [52].

The master equation method specifically considers each possible configuration of species. For a system where only H and $H_2$ are considered on the grain, possible configurations would be $(1, 0)$ with one H atom and no $H_2$ molecules on the grain, $(0, 0)$, $(0, 1)$, $(1, 1)$, $(2, 0)$, etc.. For each configuration a separate rate equation is constructed. The number of possibilities is clearly infinite, but selective cut-offs can be used to exclude the higher order terms if their probability of occurrence becomes very small. However, if the number of species expands, either by a change in physical conditions or by increasing the number of considered species in the model, the number of equations blows up rapidly. Moment equations with a cut-off up to the second-order have been suggested to make the effect less dramatic and to extend the range in which the rate equation method is applicable [53,54] and codes are available which allow this method to be used for large chemical networks in a hybrid sense in which the method is only used when the accretion limit is reached for a species [55].

Apart from stochastic methods, some modifications to the rate equations have been put forward to mimic the stochastic behaviour of the surface chemistry. Caselli et al. [56] made semi-empirical adjustments to the rates of a selection of reactions, for the case where the surface migration of atomic hydrogen is significantly faster than its accretion rate onto grains. While this method showed only limited success, the more recent modified-rate approach suggested by Garrod et al. [57] has been much more successful. In this approach, the expressions for the reactions are adjusted if the modified reaction rate is smaller than the classical rate. The modified reaction rate is determined by the accretion rate of one of the reactants multiplied by the product of the surface abundance of the other reactant and an efficiency term that takes into account the competition between diffusion across the surface for the reactants and the desorption of the reactants. Its main advantages are that it is computationally inexpensive compared with stochastic methods and that it automatically switches to the normal rate equation in the regime where those are still valid. Garrod et al. [58] have shown that the new method produces an excellent match to macroscopic Monte Carlo solutions to full gas-grain chemical systems for a range of physical conditions.

### 4.3.2.3 Methods that include layering

Species that land on an ice mantle can only react with species in the top few monolayers. Reactions deeper in the mantle either occur through bulk diffusion, which is much slower than surface diffusion, or through energetic processing such as UV photodissociation, or bombardment by energetic particles such as cosmic rays. Cuppen et al. [59] showed with a model that takes the positions of all individual species into account that the more common rate equation methods overestimate the effect of a changing gas phase composition on the grain composition by several orders of magnitude. Due to the layering of the ice, only the top layers are available for reaction and changes in the gas phase abundances are only reflected in these top layers. For simulations at constant temperature, grain models with layering show therefore many fewer fluctuations in the abundance of surface species.

The first gas-grain model that accounted for the distinction between mantle and surface was the three-phase model by Hasegawa & Herbst [60]. Here the mantle and the surface are treated as separate phases and all three phases are described by rate equations. Species can only react and desorb from the surface phase. Upon desorption the surface is replenished by species from the mantle phase and *vice versa* upon accretion. With only two solid phases (surface and mantle), they observed that the abundances at early times are preserved for a much longer period. Fayolle et al. [61] made some modifications to this model to account for bulk diffusion (or segregation), which leads to interaction between the two phases. This adjusted model is able to reproduce laboratory desorption experiments of mixed ice layers. Unfortunately, the three-phase model is not widely used, even though it has the clear advantage that it returns the correct desorption behaviour. Recently, a macroscopic Monte Carlo code has been developed that uses multiple phases for the ice mantle [62].

As mentioned in section 4.3.2.2, the microscopic implementation of the Monte Carlo method follows the position of each species on the grain. It therefore automatically accounts for layering, since only species in the direct vicinity of each other are allowed to react. It can furthermore use environmentally dependent binding energies. The main disadvantage of this method is, however, that it is computationally very expensive and it can therefore not be as easily applied for the simulation of long timescales and large chemical networks. Efforts in this direction are however on the way, including the use of massively parallel processors (Q. Chang and E. Herbst, in preparation).

### 4.3.2.4 Input parameters for surface models

All different surface models have similar input parameters. These result in (temperature dependent) hopping, desorption and reaction rates. Usually energy barriers are given from which a thermal or tunnelling rate is determined. The energy barriers, as for the gas phase, are often taken from experiments or (quantum) chemical calculations. It must be mentioned that the results of surface reaction experiments in the laboratory are not as easily converted to rates under interstellar conditions as for gas phase reactions for a variety of reasons. Firstly, the formation of molecules on the surface is the result of a combination of diffusion and reaction processes in competition with the desorption of the reactants. It is hard to disentangle the separate contributions and give barriers for the individual processes. Since all processes scale differently according to temperature and coverage, they should be disentangled before using them in an astrochemical model. Secondly, the experiments are not carried out on small particles but on large surfaces. Thirdly, the substrate, in monolayer regime experiments, is not always a good representative for a dust grain and in multilayer experiments the ice is usually much more homogeneous than an interstellar ice.

### 4.3.2.4.1 Binding energies

Desorption, which is controlled by the binding energy, is a one-step process and therefore relatively straightforward to study. Binding energies are usually determined through Temperature Programmed Desorption (TPD). In this technique, a known quantity of the species is first deposited and then the substrate is linearly heated with time while the desorption is recorded using mass spectrometry. Several of these experiments are performed with either different initial deposition quantities, deposition temperature, or heating ramps to obtain the order of the desorption process, the desorption energy, and the pre-factor. The desorption energies of a wide collection of stable species have been determined in this way. Examples are $N_2$ [63], CO [63], $O_2$ [64], $H_2O$ [65,66], and $CH_3OH$ [67]. The desorption energies have been mostly determined for the desorption of pure ices from different substrates. The differences between the different substrates are rather small and become negligible in the multilayer regime. In this regime, the molecules desorb with a (near) zeroth order rate whereas they desorb with a (near) first order rate in the monolayer regime. Since interstellar ices are not homogeneous, the desorption of mixed layers is more relevant for astrochemical modelling. However, the introduction of more species in the ice makes the desorption process immediately much more complex. First, the desorption energy can change depending on its surrounding material. Second, the dominant mantle species can prevent other species from desorbing. Collings et al. [68] showed, for instance, that molecules like CO and $CO_2$ can become trapped in an ice mantle that consists predominantly of water ice as the desorption of water occurs at much higher temperatures than for CO and $CO_2$. However, at the long timescales available in the ISM, some of these trapped species might be able to escape because of a segregation process where the two main fractions of the mantle slowly separate [69]. The model by Fayolle et al. [61], which was described earlier, is especially designed to handle this behaviour with the addition of only a few extra parameters.

### 4.3.2.4.3 Diffusion barriers

Diffusion barriers are very hard to measure and are mostly determined by quantum chemical calculations. The quality of these data relies heavily on the potential and substrate used. Experimentally, diffusion barriers can be inferred by detecting reaction products between the mobile species of interest and some immobile, sparse other reactant. Matar et al. [70] were able to determine the diffusion of H on amorphous solid water by reaction of H and $O_2$. In chemical models, often the approximation is made that the diffusion barrier is a constant fraction of the desorption energy; values of 0.3, 0.5 or 0.78 are usually taken.

### 4.3.2.4.4 Reaction barriers

Experimental studies of ice reactions can be roughly divided into two groups depending on the analysis technique and the thickness regime (sub-monolayer versus multilayer). The reactants and products can be probed mass spectrometrically. The surface is initially exposed to a small quantity of the reactants, after which the surface is heated until the products and reactants desorb and are detected via TPD. Different initial exposures and temperatures can be probed to obtain information on reaction order, etc. The main advantage of this technique is the sensitivity, which allows for sub-monolayer exposures and which is able to detect all species (masses). The method has, however, four major disadvantages: the products cannot be probed in-situ, i.e., during the atom bombardment, additional reactions during the heat-up to desorption cannot be excluded, quantifying the desorbing species is not straightforward, and some of the interesting species have equal masses.

A second method is to initially grow an ice of several monolayers and expose this ice to the atomic beam while recording reflection adsorption infrared spectra (RAIRS). In this way, the reactants and products are probed in-situ at the time and temperature that one is interested in, which is the main advantage of this technique. Quantifying the formed product is relatively simple, provided that the RAIRS is calibrated with an independent method. The main disadvantages are that not all species can be detected in this way and that the sensitivity is less than with the previous technique. Most systems have an additional quadrupole mass spectrometer (QMS) installed. So far this technique has been applied to unravel the formation of the main components of interstellar ices, *i.e.*, water, methanol, carbon dioxide, formaldehyde, and formic acid, mainly through H-atom additions to CO- and/or $O_2$-ices under conditions relevant to the interstellar medium (e.g., [70-77]). The emphasis has been on the smaller species, with ethanol being the largest product formed [78], and mostly with H-atom beams. An important reason for the latter is that purely atomic beams can never be achieved and at these low temperatures pollutants like $O_2$ or $N_2$, will stick to the surface as well; the majority of $H_2$, on the other hand, will desorb. Another important reason is that H atoms diffuse quite rapidly even at low temperatures.

As explained above, the formation of new molecules on the surface is a combination of several independent processes, especially when the stable, detectable products are the result of multiple reactions. The barriers for the individual processes are very hard to disentangle and this is only possible with additional infor-mation from, for instance, TPD experiments. For this reason, usually only a qualitative measure for the efficiency of the reaction is given. Common assessments are: (i) effectively barrierless, (ii) with a barrier that allows the reaction to proceed at 10 K, (iii) with a larger barrier so that reaction does not proceed at this temperature, (iv) thermally activated, or proceeds through quantum chemical tunnelling.

Heavier species become increasingly more difficult to form by simple atom-addition reactions. H-atom addition to acetaldehyde leads not only to ethanol, but also to the smaller organic species formaldehyde and methanol [78]. Furthermore, there is no experimental evidence that one can form longer carbon chains on the grains by simply adding carbon atoms. A different type of mechanism has been suggested that leads to the production of terrestrial-type organic molecules in star-forming regions of the interstellar medium [25]. At higher temperatures, heavier species than atoms can diffuse more readily, but they are typically not reactive because of chemical activation energy barriers. If these stable species can be converted to radicals by photons or energetic particle bombardment, rapid diffusive reactions are possible, leading to a variety of more complex species. Laboratory experiments on surface photochemistry have been undertaken for some time, but only recently under ultra-high-vacuum conditions, which are necessary to have little contamination and ice thickness comparable to interstellar ices [79].

## 4.4 Application: the molecular oxygen chemistry

The larger the molecules, the more chemical processes are involved in their synthesis and thus the more our knowledge of their chemistry becomes thinner. One would expect that the synthetic pathways of simple molecules such as molecular oxygen would be simple and by consequence well known nowadays. As we will show, however, this view is far from being correct and our understanding of the processes involved in interstellar chemistry is only improving little by little by combining modelling, laboratory experiments, and astrophysical observations.

### 4.3.1 Gas-phase synthesis

Molecular oxygen was included in chemical networks at the beginning of astrochemistry [80]. In the gas-phase, $O_2$ is a product of the neutral-neutral reaction $O + OH \rightarrow O_2 + H$, which is an exothermic, relatively fast reaction. The rate coefficient of this reaction has been measured at low temperature, using the CRESU apparatus (REFERENCE TO ANOTHER CHAPTER) by Carty et al. [81]. They found a value of 3.5 $\times 10^{-11}$ cm$^3$ s$^{-1}$ between 39 and 142 K. The OH radical is formed by the dissociative recombination of protonated water $H_3O^+$, itself formed in the following sequence of reactions:

- $H_2$ + cosmic-ray particle $\rightarrow H_2^+$ + $e^-$ + cosmic-ray particle
- $H_2^+ + H_2 \rightarrow H_3^+ + H$
- $H_3^+ + O \rightarrow OH^+ + H_2$
- $OH^+ + H_2 \rightarrow H_2O^+ + H$
- $H_2O^+ + H_2 \rightarrow H_3O^+ + H$

This scenario was first proposed by Herbst & Klemperer [80], who discussed the importance of the $H_3^+$ cation in the ISM (see also [82]).

Under dense cloud conditions (temperature around 10 K, atomic hydrogen density of a few times $10^4$ cm$^{-3}$ and no UV penetration), a pure gas-phase model will predict that oxygen and carbon will mainly form carbon monoxide. The rest of the oxygen will go into $O_2$ (see fig. 4.7). At steady state, after $10^7$ yr, the predicted $CO/O_2$ abundance ratio is approximately 3. Such a model result is strongly dependent on the assumptions concerning the elemental abundances of oxygen and carbon. The results presented here have been obtained assuming that the C/O elemental ratio is 0.4, as typically assumed based on some observations of atomic lines of carbon and oxygen in the diffuse medium [15]. We will come back later to this problem. Following this prediction, interstellar molecular oxygen has been searched for for many years.

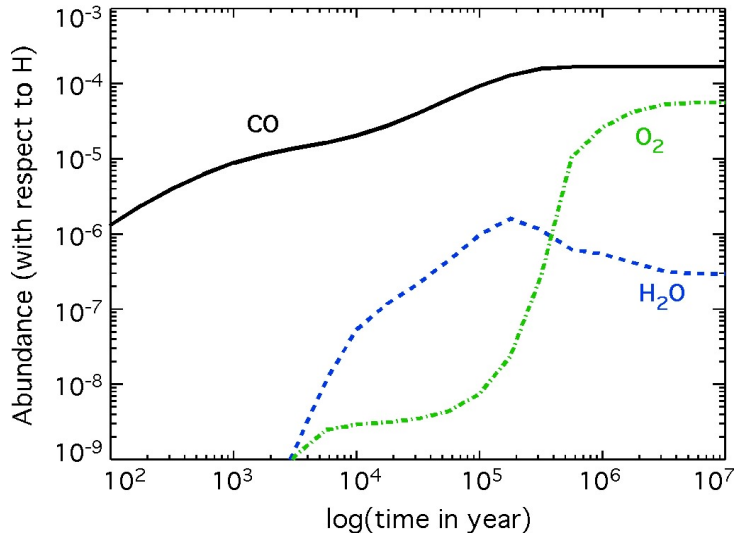

**Fig. 4.7** CO, $O_2$ and $H_2O$ abundance predicted by a gas-phase model under dense cloud conditions.

### 4.3.2 Observational constraints on $O_2$ abundance

Since the 1980's, $O_2$ has been looked for in the interstellar medium using both ground-based and space telescopes (see [83], and references therein). First, analyses of data from the SWAS satellite gave an upper limit of about $10^{-6}$ in dense cold clouds [84]. After the first detection claim by Pagani et al. [83], Larsson et al. [85] announced the detection of $O_2$ by re-analysing data from the ODIN satellite and published a beam-diluted abundance of $5 \times 10^{-8}$ relative to $H_2$. Using ground based observations of $O^{16}O^{18}$ and $C^{18}O$ lines, Liseau et al. [86] argued that the emitting region may be much smaller than the beam of ODIN and thus the $O_2$ abundance could be larger by one or two orders of magnitude. More recently, the Herschel satellite found three magnetic dipole rotational transitions of $O_2$ towards the $H_2$ Peak 1 position of vibrationally excited molecular hydrogen in Orion KL [87]. The fractional abundance of $O_2$ relative to $H_2$ was found to be $(0.3\text{-}7.3) \times 10^{-6}$. The authors suggested the source of the $O_2$ to be either thermal evaporation from warm dust or the passage of a C-shock. Why the molecule is only found in such an unusual source remains a mystery.

### 4.3.3 Gas-grain interactions

Based on these new observational constraints, models have been altered in order to decrease the predicted abundance of gas-phase $O_2$. Problems in the rate coefficients of gas-phase reactions have been looked for. The very low temperature rate coefficient of the O + OH reaction has been investigated theoretically, but Quan et al. [88] showed that only a seriously unrealistic decrease of this rate coefficient would really impact the predicted $O_2$ abundance.

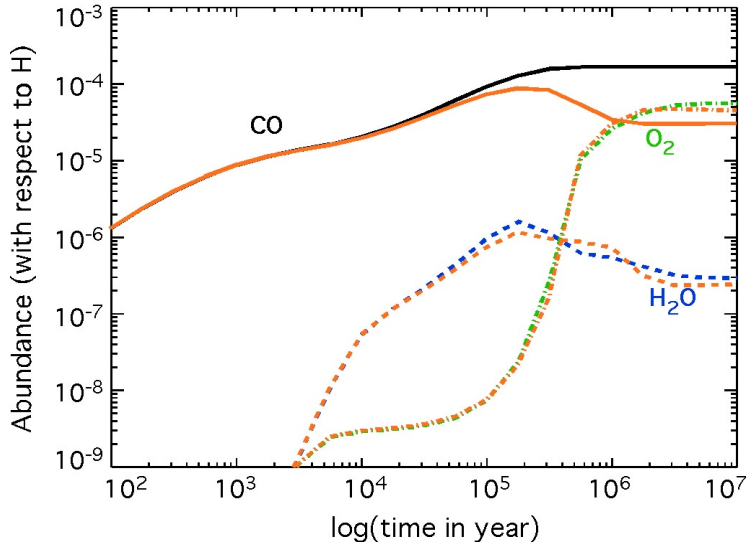

**Fig. 4.8** CO, O$_2$ and H$_2$O abundances predicted by a gas-phase model under dense cloud conditions (same as Fig. 4.7) and by a gas-phase model including sticking and evaporation of species from grain surfaces (orange lines).

Other efforts have been made to investigate the interaction of gas-phase O$_2$ with interstellar grains (see for example [89-91]). Once O$_2$ is formed in the gas phase, the molecule sticks significantly on the surface of interstellar grains in approximately $10^5$ yr, decreasing the gas-phase abundance of O$_2$. Desorption induced by cosmic rays, however, is efficient enough to desorb the molecule and the gas-phase abundance remains almost unchanged if only sticking and evaporation are considered. This result is shown in fig. 4.8, where we overlay the abundances predicted by a pure gas-phase model with a model including sticking of gas-phase species on grain surfaces and desorption of these species from the surfaces (section 4.3.2). Once they are on the grain surfaces, however, molecules can react with other species present.

### 3.4 Grain surface reactions

At temperatures as low as 10 K, atoms can move efficiently and react with other molecules accreting from the gas-phase. Following this idea, Tielens & Hagen [90] suggested that O$_2$ on the surface can be successively hydrogenated by reactions with atomic hydrogen to give H$_2$O$_2$ and then produce water, again by a further reaction with atomic hydrogen. Recently, this surface reaction scheme has experimentally been proven to be even more efficient than assumed in the models of Roberts & Herbst [70,73,74]. This formation path is, however, only one of several possibilities to transform oxygen into water on surfaces. If the direct hydrogenation of $O_2$ on the surface is removed, other reactions such as the hydrogenation of ozone or the reaction between OH and $H_2$, would take its place to remove the oxygen (whatever its form) and transform it into $H_2O$. Whatever the exact scenario, the predicted abundance of gas-phase $O_2$ is then strongly decreased at late times (see fig 4.9). A peak of $O_2$ up to about $5 \times 10^{-6}$ (compared to H) however remains at typical ages of dense clouds (a few x $10^5$ yr).

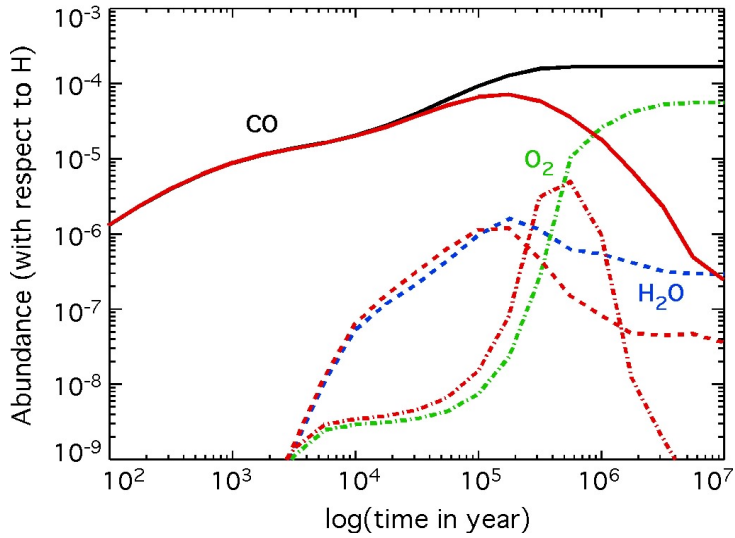

**Fig. 4.9** CO, $O_2$ and $H_2O$ abundance predicted by a gas-phase model under dense cloud conditions (same as Fig. 4.7) and by a full gas-grain model (red lines).

Considering the difficulty to observe $O_2$ in quiescent dense regions, unlike that observed by Goldsmith et al. [87], and the lack of spatial resolution of the current observations, it is difficult to make any conclusions on the agreement between models of cold cores and the observational constraints. One may think that the scarcity of the detections despite the number of target sources (perhaps covering a large range of cloud ages) is an indication that $O_2$ is even less abundant than predicted by these gas-grain models. If so, work still has to be done to solve this problem but probably not with a purely chemical point of view. Chemical model predictions do indeed depend on other parameters than chemical reactions and rate coefficients. Elemental abundances are among the most important ones and re-main quite uncertain (see section 4.2). Observations in diffuse clouds show that the elemental C/O ratio does not vary much with the line of sight and stays around 0.4 [15]. It has been recently proposed by Jenkins [8], however, that oxygen could be slightly more depleted than carbon in denser sources, a possibility also discussed by Whittet [92]. As a consequence, the C/O elemental ratio in the gas phase of dense regions could be larger than what is typically assumed. This idea was used by Hincelin et al. [93] to explain the low $O_2$ abundance in dense clouds. Using a C/O ratio greater than unity with a full treatment of the gas-grain chemistry would produce an abundance of gas-phase $O_2$ smaller than $5x10^{-8}$ (compared to the total atomic hydrogen density $n_H$) at all times and could explain the observations directed at cold cores, without changing strongly the abundances predicted for other species.

## 4.3.5 Importance of nitrogen chemistry for $O_2$

Current networks for the ISM can contain more than six thousand reactions, including both gas-phase and grain-surface processes. As discussed in Sec. 2.1.3, reduction of chemical networks can be done to allow the coupling with a dynamical model that otherwise would be very time consuming [94,95]. Such reductions have to be performed very carefully and can only be accomplished for a specific condition. The chemistry of $O_2$ provides a good example of this fact. With a C/O elemental abundance of 1.2, instead of 0.5, nitrogen chemistry becomes important for the $O_2$ abundance although no N-bearing species are involved in its synthesis [93]. Under dense cloud conditions and with a gas-grain model (and an elemental C/O ratio of 1.2), the CN molecule is abundant and reacts with $O_2$ to form either O + OCN or CO + NO. The total rate coefficient of this reaction has been measured by Sims et al. [96] between 13 and 295 K, and the product branching ratios by Feng & Hershberger [97] so that the temperature dependent rate coefficients are:

- CN + $O_2$ → O + OCN     $k_1$ (T) = $1.992x10^{-11}$ (T/300)$^{-0.63}$ cm$^3$ s$^{-1}$ ,
- CN + $O_2$ → CO + NO     $k_2$ (T) = $4.98x10^{-12}$ (T/300)$^{-0.63}$ cm$^3$ s$^{-1}$ .

The rates of these reactions also depend on the abundance of CN, a molecule significantly destroyed by the neutral-neutral reaction with atomic nitrogen. The rate coefficient of the reaction N + CN has recently been revised. The value from the osu_01_2009 database [98] for N + CN → $N_2$ + C is $3x10^{-10}$ cm$^3$ s$^{-1}$, and is temperature independent. Some experimental measurements between 56 and 296 K have shown a decrease of the rate coefficient at low temperature according to the expression $8.8x10^{-11}$ (T/300)$^{0.42}$ cm$^3$ s$^{-1}$, which gives a rate coefficient of 2 x10$^{-11}$ cm$^3$ s$^{-1}$ at 10 K by extrapolation [99]. Such a decrease of this rate coefficient has a strong effect on the abundance of CN, which is then increased, and as a consequence on $O_2$, which is decreased. This result is shown in fig. 4.10, where the $O_2$ and CN gas-phase abundances are displayed, for the same dense cloud conditions as in the previous figures in this section, and for two N + CN rate coefficients at 10 K: the larger one from the osu database and the smaller one as extrapolated from the recent experiments (see also [93]). It can be seen in the figure that the $O_2$ abundance is now reduced to a maximum of $10^{-7}$ with the smaller rate coefficient, in better agreement with observations of cold cores. With a C/O elemental ratio of 0.5, the $O_2$ gas-phase peak abundance remains larger than $10^{-6}$ and is insensitive to the CN abundance.

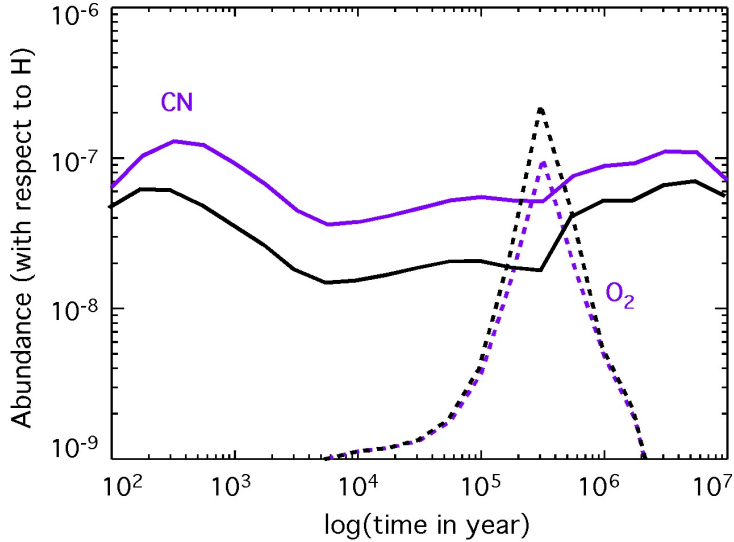

**Fig 4.10** Gas-phase CN and $O_2$ abundances predicted by a gas-grain chemical model under dense cloud conditions (same as Fig. 4.7) and a C/O elemental ratio of 1.2, as a function of time. Black and purple curves were obtained for a C + CN rate coefficient of $3 \times 10^{-10}$ cm$^3$ s$^{-1}$ and $2 \times 10^{-11}$ cm$^3$ s$^{-1}$ at 10 K respectively.

## 4.4. Summary and concluding remarks

Chemical models are important for the analysis of observations of molecules in the interstellar medium and to make predictions for molecules that have not or cannot be directly observed. Modern models are able to compute the evolution of the chemical composition of a mixture of gas and dust taking into account a large number of processes including bimolecular gas phase reactions, interactions with cosmic-ray particles and UV (and X-ray) photons, interactions with interstellar grains, and grain-surface reactions. To account for all these processes, chemical networks contain thousands of reactions, each characterised by a rate coefficient associated with an uncertainty. Uncertainties in other model parameters, such as elemental abundances, temperature, density, etc., have also to be taken into account while discussing the accuracy of these models; in particular, when comparing with observations.

Depending on the astrophysical objects studied, the geometry and the physical dynamics may have to be included, making these models more and more complicated. With the start-up of a new and powerful interferometric telescope labelled ALMA, an acronym for Atacama Large Millimeter Array, with its high angular and spectroscopic resolution, and high sensitivity, the predictions of chemical models will be compared with much higher quality data. As an example, ALMA observations at high spatial resolution will resolve smaller structures in protostellar envelopes, which will rule out the use of spherical symmetry for the chemistry (see also Herbst [100]). For these reasons, these models have to be improved in the future. In addition to the coupling with better constrained physical structures, the chemistry itself has to be improved. Methods, such as sensitivity analysis, have been developed to identify the key processes in the gas phase and quantify the model accuracy. An interactive user-friendly database for gas-phase processes has even been created with the aim of improving the visibility of available data. The name of the database is KIDA, which is an acronym for KInetic Database for Astrochemistry;[6] this database is updated regularly. The most uncertain part of astrochemical models is probably all the processes related to surface reactions. Sensitivity analyses, as described in this chapter, are in principle applicable to the grain surface processes based on the rate equation method, but the number of parameters to study would be very large and drawing definitive conclusions not so easy. In addition, it is the nature of the processes themselves, rather than the parameters, that is more uncertain. Much progress has been made with experiments of surface chemistry but it remains difficult to use these experiments to improve the models because, unlike gas-phase experiments, the results of surface and ice experiments are not so easily converted from the laboratory to the very different conditions in the low-density ISM. The first step towards improvement could be the construction of a database to centralise all the information for surface reactions. The existence of this database might limit the multiplicity of gas-grain models based on different networks.

**Acknowledgments**  V. W.'s research is supported by the French INSU/CNRS program PCMI, the Observatoire Aquitain des Sciences de l'Univers, and the Agence Nationale de Recherche (ANR-JC08–311018: EMA:INC). H.C. thanks the European Research Council (ERC-2010-StG, Grant Agreement no. 259510-KISMOL) and the Netherlands Organisation for Scientific Research (NWO) (VIDI) for financial support. E. H. acknowledges the support of the NSF (US) for his research program in astrochemistry and the support of NASA for his program in exobiology.

---

[6] http://kida.obs.u-bordeaux1.fr/